
%
%
%
\def\gtaprx {\lower .1ex\hbox{\rlap{\raise .6ex\hbox{\hskip .3ex
	{\ifmmode{\scriptscriptstyle >}\else
		{$\scriptscriptstyle >$}\fi}}}
	\kern -.4ex{\ifmmode{\scriptscriptstyle \sim}\else
		{$\scriptscriptstyle\sim$}\fi}}}
\def\ltaprx {\lower .1ex\hbox{\rlap{\raise .6ex\hbox{\hskip .3ex
	{\ifmmode{\scriptscriptstyle <}\else
		{$\scriptscriptstyle <$}\fi}}}
	\kern -.4ex{\ifmmode{\scriptscriptstyle \sim}\else
		{$\scriptscriptstyle\sim$}\fi}}}
\def \yskip{\penalty-50\vskip3pt plus3pt minus2pt}
\def \yyskip{\penalty-100\vskip6pt plus6pt minus4pt}
\def\pp{\par\noindent\hangindent 0.4in \hangafter 1}
\def\singlespace{%
   \lineskip            0.15ex
   \baselineskip        3.0ex
   \lineskiplimit         0ex
   \parskip             0.6ex plus .30ex minus .15ex
   }%
\def\oneonehalfspace{%
   \lineskip             .20ex
   \baselineskip        4.0ex
   \lineskiplimit         0ex
   \parskip             0.80ex plus .40ex minus .20ex
   }%
\oneonehalfspace
\magnification 1090
\baselineskip 27pt
\vskip .5 truein
\centerline {\bf N-body Simulations of Star-Disc Captures in Globular Clusters}
\vskip .2truein
\centerline {\bf S. D. Murray\footnote{1}{{\rm Postal Address: Dept. of
Astronomy, 601 Campbell Hall, Univ. of California, Berkeley, CA 94720, U.S.A.}}
and C. J. Clarke}
\vskip .2truein
\centerline {\bf Institute of Astronomy,}
\centerline {\bf Madingley Road,}
\centerline {\bf CAMBRIDGE, CB3 OHA, U.K.}
\vskip 1truein
\noindent {\bf ABSTRACT:}  The presence of protostellar disks can greatly
increase the dissipation during close stellar encounters, leading to the
formation of a significant population of binaries during the initial collapse
and virialization of a cluster.  We have used N-body simulations of collapsing
globular clusters to find the major factors that determine the efficiency of
binary formation through star-disk captures.  This work serves the dual purpose
of verifying the results of earlier analytic work as well as examining
parameters not testable by that work.  As in the earlier work,
typical binary fractions of a few percent are found.  For the parameters
studied, the results are found to depend remarkably little upon disk evolution,
the mass distribution of the stars, or their spatial distribution, though
distributions in which the stars are highly clumped yield binary fractions
larger by a factor of a few.  The direct N-body integrations limit the models
to relatively small values of N.  Semiempirical relations are derived,
however, which allow the results to be extrapolated to values of N appropriate
to globular clusters.

\vfill\eject

\noindent {\bf 1 INTRODUCTION}
\par A number of recent studies suggest that globular clusters contain a
significant population of binaries (Pryor, Latham \& Hazen 1988;
Pryor et al 1991; Romani \& Weinberg 1991). It is  widely believed that the
majority of these binaries must be primordial (i.e. formed at the epoch
of cluster formation) since negligibly few would  have formed subsequently
through three-body capture processes (Hills 1976), and only the closest
binaries can be ascribed to subsequent tidal capture events (Fabian, Pringle
\& Rees 1974).  The presence of a number of primordial binaries in globular
clusters has also long been suspected on theoretical grounds, since they
play an important role in averting the onset of gravothermal catastrophe
and cluster core collapse (Goodman 1989; Goodman \& Hut 1989; McMillan,
Hut \& Makino 1990, 1991; Gao et al 1991).
\par In this paper, we consider in some detail a
proposed mechanism for the formation of primordial binaries in globular
clusters, namely that of captures due to dissipation by protostellar disks
during close encounters in an initial phase of
cluster violent relaxation. This possibility was examined by Murray, Clarke
\& Pringle (1992, henceforth MCP) using an idealised analytical model which
predicted binary fractions of a few per cent--a number broadly compatible
with the results of binary surveys in globular clusters. Such an analytical
model is however only applicable to the case of the homologous collapse of
an initially uniform density sphere, thus raising the suspicion that its
results were heavily reliant on this (possibly unrealistic) initial condition.
In order to remedy this uncertainty, therefore, we have undertaken a number of
N-body simulations of the violent relaxation of stellar clusters, with the
aim of discovering how the binary capture rate is affected by the initial
global density profile, the degree of homogeneity and the spectrum of
stellar masses involved.
\par Before setting out the star-disc binary formation mechanism in more
detail, it is first necessary to describe the context in which such a mechanism
would operate in the early stages of globular cluster formation. We here
follow a number of previous authors in assuming that stars form in globular
clusters with a velocity dispersion that is substantially sub-virial, and
that virialisation then proceeds on a dynamical time scale through the process
of violent relaxation (e.g. Aarseth, Lin \&
Papaloizou 1988, henceforth ALP).
Such a picture is clearly at odds with the way in
which star clusters are currently observed to form in the Galactic disc, since
here star formation appears to proceed in an ensemble of cloud clumps whose
motions are already substantially virialised in the potential of the parent
cloud. In globular clusters, however,
a much more rapid (dynamical) process of fragmentation and star
formation is indicated by a number of considerations.  The narrow red giant
branches of globular clusters are used to infer internal metallicity
spreads $\Delta[$Fe/H$]$ ranging from 0.1 for metal poor clusters to 0.01 for
metal rich clusters (Sandage \& Katem 1977, 1982; Cohen 1979; Richer \&
Fahlman 1984; and Bolte 1987ab).  Such small metallicity spreads imply that
star formation must have occurred on a time scale less than a cluster dynamical
time so as to avoid self-enrichment (Murray \& Lin 1989).  A similar limit
may result from the narrow observed widths of the red giant branches of
massive, young clusters in the Magellanic Clouds (Elson 1991).  A dynamical
constraint results from the fact that slower, less
efficient star formation could lead to the clusters becoming unbound due
to the loss of gas resulting from ionization-shock fronts caused by the first
massive stars to form (Tenorio-Tagle et al. 1986; Lada, Margulis, \&
Dearborn 1984).  If star formation indeed occurs on a dynamical
time scale, this suggests that some
external trigger (such as a cloud-cloud collision)
must have brought about the sudden fragmentation of a cloud previously in
approximate
hydrostatic equilibrium (Murray \& Lin 1989). In this case,
the stars would fragment out
of the background on essentially radially infalling orbits, causing the
collapse of the entire system. The formation of a singularity is however
avoided due to
the velocity dispersion generated by global gravitational instabilities
during the collapse (ALP).
Consequently, the system `bounces' at a finite radius ($N^{-1/3}$ of the
initial
cluster radius for an initially homogeneous sphere containing N mass points)
and rebounds into a state of approximate virial equilibrium.
\par It is in this context that we consider the role of star-disc interactions
in the formation of binaries. Here, the idea is that two-body encounters
between stars are dissipative if they involve a star's  passage close to (or
through) the protostellar disc around another star (Pringle 1989; Larson 1990).
Clarke \& Pringle (1991a) (henceforth CP) however argued
against such a process as an important source of binaries in the case of large
N {\it virialised} clusters, owing to the prevalence of fast (non-capturing)
encounters that would act so as to destroy the protostellar discs. The present
situation is different, however, since the initial stellar velocity dispersion
is very low: the binary formation rate changes during the collapse as a result
of a trade-off between the competing effects of growing velocity dispersion
and rising density. MCP quantified these effects using the analytical
estimates of ALP for the growth of density and velocity dispersion during the
collapse of a uniform sphere and applying a simple criterion for the
dissipative energy loss during star-disc encounters. This exercise yielded
a binary formation rate that rose during the collapse, peaking at the bounce,
at which point the calculation was terminated due to the breakdown of the
analytical estimates in the non-linear regime. MCP however anticipated that
there would be little subsequent binary formation since the high velocity
dispersion during the bounce would destroy most remaining discs.

\par In this paper, we address the same problem using an N-body code. This
has a number of advantages over the method described above, being more flexible
in terms of the initial conditions that it can handle and allowing the
calculation to be pursued into the  non-linear regime--i.e. through and beyond
the bounce phase. We therefore consider both uniform and isothermal initial
density profiles, these corresponding respectively to the limits in which the
proto-globular cloud was predominantly confined by an external medium and by
self-gravity (we do not, however, consider the deviations from spherical
symmetry induced by the external trigger for fragmentation and star formation:
see Boily, Clarke \& Murray 1992). We also consider the case of an initially
clumpy mass distribution (motivated by the appearance of young globular
clusters in the Magellanic clouds (Elson 1991)) and investigate the effects
both of including a spectrum of stellar masses and of varying the initial
stellar
velocity dispersion. An obvious disadvantage of N-body codes, compared with
the analytical method of MCP, is however the steeply rising computational
cost with increasing values of N. Since it is clearly unfeasible to run
simulations with N of order the number of stars in a globular cluster, and
since a variety of effects mean that the scaling of binary formation rate with
N is not clear from first principles, we have been forced to experiment with
different values of N.  When (in the uniform density case) a value of N is
attained ($>N_{min}$) for which the results agree with the analytical estimates
for that N , the analytical
results can then be used  to extrapolate into the high N regime appropriate to
globular clusters. The comparison of differing initial conditions is then made
for models with $N=N_{min}$. In practice, we find that this implies
simulations involving 2000 particles.

The structure of the paper is as follows. In \S\ 2 we set out the
numerical method and the range of models that we explore. In \S\ 3 we
briefly describe the results of these models. Section 4 contains a detailed
discussion of the results; where possible we have contrasted the results with
the analytical model of MCP, and present semi-empirical expressions for the
scaling of binary yields with disc and cluster properties. Section 5
briefly presents the conclusions.

\yyskip
\noindent{\bf 2. THE MODELS}
\yyskip

\noindent{\bf 2.1. Numerical Method}
\yyskip

The most straightforward means by which the formation of binaries can
be studied is by the use of direct N-body integration of the orbits and
interactions of the stars of a collapsing cluster of protostars.  For
long term integrations, the method is limited to $N\ltaprx10^3$ (Aarseth 1985).
In principle, larger $N$, up to $\sim10^4$, could be used in the present
study, for which the clusters need only be followed for a few dynamical
times.  So as to make the most efficient use of computation time, we have
in practice limited most of our models to 2000 stars, so as to be able to
investigate the effects of changing several parameters. To ensure that the
models do adequately represent the evolution of larger $N$ systems, we have run
some models with as many as $2\times10^4$ stars (see below).

The program used is described in Aarseth (1985).  It employs a direct
integration of the motion of the stars, with separate integration time steps
for each star.  To further improve computational efficiency, the contribution
to the force on each star is divided into two components, one from close
neighbors, and the other from more distant particles, with the latter being
updated at longer intervals.  The effects of close encounters between stars are
weakened by the use of a softened potential, in which the interparticle force
varies as
$$F_i\propto{{x_i}\over{\left(r^2 + \epsilon^2\right)^{3/2}}},\eqno{(2.1.1)}$$
where the index $i$ indicates the component, $r$ is the interparticle
separation, and $\epsilon$ is the constant softening parameter.  To ensure
that softening does not affect the capture rates, we set $\epsilon\ll R_d$,
where $R_d$ is the disk radius, in all results below.

The energy loss due to encounters with stellar disks is treated impulsively.
During the integration of a star's motion, the nearest neighbour distance
is calculated.  If this distance is less than the sum of the disk radii, then,
when
the stars reach periastron, their kinetic energy is decreased by an amount
equal to the total kinetic energy of the stellar disks at periastron (CP), or
$$\Delta E={1\over2}\left(M_{d1}V_1^2 + M_{d2}V_2^2\right),\eqno{(2.1.2)}$$
where $M_{di}$ and $V_i$ are the disk masses and relative speeds in the
center of mass frame of particles 1 and 2, respectively.  We note that the
method of treating encounters does not account for the relative orientations
of the disks, does not allow for (rare) three-body interactions, and assumes
that the disks are entirely destroyed during the encounter, thus ignoring
the effects of further interactions.  The  assumption that the disks are
completely disrupted should give an upper limit to the energy loss
per encounter (CP), but also implies that each disk can be involved in only
one encounter, and is in this latter sense a pessimistic assumption. We do
not feel that further experimentation with the interaction prescription is
warranted at this stage, pending further detailed work on the nature of
star-disc interactions (Clarke \& Pringle in preparation).
\par The N-body integration method described above has been tested against the
numerical method used by MCP, for systems
with $10^4$ stars which were initially distributed with uniform density, and
whose disk radii and masses were held constant with time.  The number of
binaries, $N_{bin}$, found by the two methods agreed to within 10\% at the
time that the analytical calculation had to be abandoned due to the
perturbations becoming non-linear. Such excellent agreement helps to confirm
both the simplified analysis of the previous work, as well as the lack of any
affect of $\epsilon$ upon the results.

\yyskip
\noindent{\bf 2.2. Models}
\yyskip

The advantage of the current method is that it allows us to examine a more
realistic range of parameters than could be studied with the previous analytic
method.  Most important are the effects of varying the number of stars, their
initial kinetic energy, the spatial distribution of the stars, and the effects
of a distribution of stellar (and disk) masses and radii.  The last three could
not be examined using the method employed by MCP.

The parameters of the models used are summarized in Table~1.  Listed are:
the model number; the number of stars, $N$; the stellar density distribution,
$\rho_\ast(R)$ where $R$ is the radius within the cluster (see below); the
initial ratio of kinetic energy to potential energy of the cluster, $Q$, (0.5
for virial equilibrium); whether or not disk evolution is included; the initial
disk radii; and the
initial mass function of the stars. Also listed in each case are the number
of binaries formed ($N_{bin}$) and the time $(t/\tau_c)$
(where $\tau_c$ is the cluster crossing time) at which
the number of binaries is evaluated. For small $N$ models ($\ltaprx 2000$)
the binary yield is normally evaluated after two crossing times, whereas at
large $N$ we have been forced to terminate our calculations earlier, either
due to the computational expense of running for longer times or because of
the accumulation of energy errors in the N-body integration to an unacceptable
degree. In all cases we estimate that the incompleteness of our binary yields
is $\ltaprx 25\%$.

Models 1-16 examine the role of cluster properties.  Each of these
models uses
equal mass stars, with $M_\ast=1$~M$_\odot$ assumed.  The disk radii and
masses, $R_d=10^{-3}$~pc, and $M_d=M_\ast/2$, respectively, are held
constant in time: we discuss what values of $R_d$ might be expected in
practice in \S\ 4.4.  The half-mass radius of each model is
$r_h\approx2$~pc, with
some variation due to the random placement of the stars. The number of stars
used in these models varies from 50 to $10^4$.

Given the uncertainties in the form of $\rho_\ast(R)$ to be expected following
star formation in clusters, three extremes have been tested.  If star formation
occurred within initially pressure-bound clouds, then the resulting stellar
distribution might be expected to be fairly uniform, reflecting the gas
density (models 1-7)  Alternatively, in initially self-gravitating clouds,
more
centrally condensed density distributions are expected.  A singular isothermal
sphere represents one extreme distribution for a hydrostatic, self-gravitating
cloud, and so we adopt $\rho_\ast\propto1/R^2$ in models~8 to 13.

Star formation may not be expected to follow a smooth density law, but may
occur in clumps.  This is observed in molecular clouds today (Shu, Adams, \&
Lizano 1985), and is also predicted in globular clusters if star formation
occurs as the result of thermal instability (Murray \& Lin 1989), in which case
stars form in regions where the gas was initially overdense relative to the
background.  To test this, the stars in model~14 are initially distributed
randomly within ten subclumps of half-mass radius 0.4~pc, which themselves are
distributed randomly throughout the cluster radius.

The final cluster parameter of interest is the kinetic energy of the stars.
To test its effect, the initial value of $Q$ is varied between 0.01 and
0.05.

Models 17-19 examine the dependence on disc parameters. In model 17 the
disc radii are increased by a factor five compared with previous models
(note that this is identical to decreasing the cluster radius by the same
factor). In models 18 and 19 we examine the effect of viscous evolution
of the discs for two (uniform and isothermal) models with
$N=2000$, and $Q=0.01$.  The disk evolution is approximated as in Lin
\& Pringle (1990), such that the disk radii and masses vary as
$${{R_d}\over{R_{d0}}}=\left({{M_{d0}}\over{M_d}}\right)^2,\eqno{(2.2.1)}$$
and
$$M_d=M_{d0}\left[1+5\left({t\over{\tau_{\nu0}}}\right)\right]^{-1/5},
\eqno{(2.2.2)}$$
where
$$\tau_\nu=\left({{R_d^3}\over{GM_\ast}}\right)^{1/2}\left({{M_\ast}\over
{M_d}}\right)^2\eta^{-1}\eqno{(2.2.3)}$$
is the viscous time scale, $\eta$ is an adjustable parameter, and $M_\ast$
is the total mass of the disk and central star.  The
models discussed below all assume $\eta=0.001$:  for the disk and cluster
parameters given above, this gives $\tau_{\nu0}=1.9$~Myr, comparable to
the initial free-fall time $\tau_{ff}=1.5$~Myr for the uniform density clusters
with $N=2000$.

The potential role of variations in stellar and disc properties is  examined in
models~20-22.  In each, the stellar distribution is assumed to follow a
Salpeter mass function, with stellar masses in the range 0.5-2.5~M$_\odot$,
giving a mean stellar mass of $\langle M\rangle=1$~M$_\odot$. The variation of
the disk masses and radii with stellar mass are uncertain, and will depend upon
the early evolution of protostellar fragments prior to star formation (see
discussion in \S\ 4.5). For simplicity, we assume a constant
disc to star mass ratio and take
$M_{d0}=0.5M_\ast$ for all stars.  We also assume $R_{d0} \propto M_*$
(with $R_{d0} = 10^{-3}$ pc for $M_* = \langle M \rangle$) but
discuss this assumption criticially in \S\ 4.5 below.

\yyskip
\noindent{\bf 3. RESULTS}
\yyskip

\noindent{\bf 3.1. Variation with N (Uniform Models)}
\yyskip

Figure~1  shows $N_{bin}$ vs. time, and the resulting distribution of
semimajor axes, $a$, for models~1-5.  The time at which the
system reaches maximum compression before re-expanding and virializing is
approximately one initial free-fall time of the clusters, or approximately
$3\ N_{500}^{-1/2}$~Myr, where $N_{500}=N/500$. Also shown are the results
of the analytical calculation (MCP) at the time that the perturbations
become non-linear.  Figure 2 illustrates the onset
of destructive star-disc collisions during the bounce for $N=10^4$.  It also
demonstrates that the rate of binary formation has, indeed, slowed down
greatly by the end of the simulation, which is not apparent from Figure~1a.

The distribution of $a$ is similar in each of the models, so that only the two
extreme cases are shown in Figure~1b.  From the figure, it can be seen that the
distribution is similar to that found by Clarke \& Pringle (1991b) for small
$N$ systems, and predicted by MCP for large $N$ systems.  The distribution in
$\log a$ has a peak near $a=R_d$, and for $a<R_d$, the number of binaries with
semimajor axes less than a given $a$ varies approximately as
$N_{bin}(<a)\propto a$.

\yyskip
\noindent{\bf 3.2. Variation with Density Law and Initial Kinetic Energy}
\yyskip

The solid and dotted lines in Figure~3 show the evolution of $N_{bin}$ with
time for Models 3 and 6, two uniform clusters containing 2000 stars with
different values
of the initial kinetic energy ($Q=0.05$ and $0.01$ respectively). Again the
crosses mark the corresponding analytical results when the perturbations
go non-linear. At this stage, it is evident that the binary yield is rather
sensitive to $Q$ (differing by approximately 50\% after 2.5~Myr); the similar
post-bounce yields
however diminish the contrast in the total number of binaries. For higher
$N$ values, the shutting off of binary production after the bounce leads to
the marked Q-dependence being preserved in the total yield of binaries
(model 4 cf model 7, Table 1).
\par Also shown in Figure 3 (short-dashed and long-dashed lines) are the
corresponding plots for Models 10 and 12, $N=2000$ models in which the initial
density profile is that of an isothermal sphere. It is immediately apparent
that whereas the time-dependence of the binary capture rate is quite different
as compared with the uniform case (note the absence of a well defined `bounce'
phase of peak binary production in the isothermal case) the over-all binary
yield is changed by less than a factor two. The isothermal case is considerably
less sensitive to $Q$, at this $N$, implying that the collapse process is more
efficient in erasing any trace of the initial conditions than in the uniform
case. At larger $N$, however, this $Q$ dependence of binary production is
better preserved in the isothermal case as well (Figure 4).
\yyskip
\noindent{\bf 3.3 Clumpy Initial Conditions}
\yyskip

Figure~5  shows the evolution of Models 6 and 14,  both of which have
$N=2000$
and $Q=0.01$.  In Model 6, the stars are distributed uniformly, whereas
in model 14 they follow a clumpy distribution as described in \S~2.2.
As expected, the clumps collapse ahead of the overall infall of the cluster,
and the binary yield is increased by a factor of about two relative to the
uniform case. We have also run a model (17) corresponding to a single
clump in model 14, the similar binary fraction obtained in this case
confirming that model 14 evolves like an ensemble of independent clumps
over several cluster crossing times. We have also included in Table I a
selection of low $N$ models (1, 15-17) that can be used in order to
estimate binary fractions from clumpy initial conditions.

\yyskip
\noindent{\bf 3.4. The Role of Disk Evolution}
\yyskip

Figure~6a compares the evolution of models
6 and 18,  both of which are uniform density clusters with $N=2000$, and
$Q=0.01$.  In
model~6, the disk properties do not evolve with time, while in model~18
$R_d$
and $M_d$ evolve as described in \S\ 2.2 above.  At early times
($\ltaprx1.4$~Myr), the increase in $R_d$ due to viscous evolution increases
the capture rate relative to the unevolving system.  At
later times, however, the capture rates are approximately equal in the two
clusters, an effect that can be traced to the steep increase in $N_{loss}$ at
late times in the viscously evolving case (dotted and long dashed curves in
Figure 6a)).
(By the last times shown, $N_{loss}=297$ in model~18, as compared to
111 in model~6.)  The primary effect of disk evolution is thus to increase the
overall encounter rate significantly, while increasing the rate of captures
only slightly.

Evolution of the disk radii also affects the distribution of $a$, as shown
in Figure~6b.  Because more captures occur when $R_d$ has a greater value,
the peak in the distribution is shifted to a larger $a$ by about a factor
of two, corresponding to the value of $R_d$ near the time of maximum stellar
density, with a corresponding increase in the width of the distribution.

The change in $N_{bin}$ is even less striking for the case of centrally
condensed systems, as is apparent in Figure~7a, which shows the evolution of
models~12 and 19.  The capture rate in these systems decreases slowly with
time, in contrast to the uniform density models which are dominated by
captures during the bounce phase.  As a result, in the viscous case, many
encounters occur before the disks evolve significantly. The increase in the
encounter rate at late times relative to the case in which the disks do not
evolve occurs at the expense of an increase in $N_{loss}$. Rather than
resulting in a shift in the peak of the distribution in $a$, the effect is to
broaden the distribution, which is now almost uniform in the range
$1.8\ltaprx\log a\ltaprx2.8$, as shown in Figure~7b.

\yyskip
\noindent{\bf 3.5. The Role of the IMF}
\yyskip

Models 20-22 are equivalent to models 18, 6 and 19, except that the
stellar mass distribution follows a Salpeter function over a factor five
in mass: in each case, the resultant binary yield differs from the
equal mass case by $\ltaprx 50\%$.

Figure ~8 demonstrates the mass dependence of the binary formation process:
in  Figure 8a the fraction of stars of a given mass that are contained
in binaries is plotted as a function of stellar mass (models 20 and 22),
illustrating the preferential incorporation of higher mass stars into
binaries. This same effect is also illustrated in Figures~8b and 8c which
plot the mass ratio $M_2/M_1$ as a function of $M_2$ for binaries in which
$M_1$ lies in the range $0.7-1.3 M_\odot$ (in the event that both binary
members lie in this range, $M_1$ is taken to be the star with mass closest
to $1 M_\odot$). In each case the distribution resulting from random
pairing from the mass function is also illustrated: again, the preference
for high mass companions is demonstrated, particularly for the
isothermal case.

\yyskip
\noindent{\bf 4. DISCUSSION}
\yyskip

\par
We now discuss the behaviour of the models described above and attempt to
understand these results by comparing them with the scalings suggested by
analytical estimates. In this way, we are able, in certain regimes, to
propose semi-empirical expressions for the resultant binary fraction as a
function of model parameters. Before considering various regimes in detail,
however, we first lay out the analytical dependences that govern the rate
of binary formation by star-disc capture. For the star-disc capture
prescription used here (i.e. in which the relative orbital energy of two
stars is reduced by an amount proportional to the disc's orbital kinetic
energy, if periastron is less than $R_d$, and is unchanged if periastron is
greater than $R_d$, independent of orbital inclination) the appropriate
capture rates as a function of disc and local cluster variables are given by
equations (2.2.8) and (2.2.9) in CP (note that, strictly speaking, such a
formulation is only appropriate to the case of a Gaussian local velocity
dispersion, a
condition that is not necessarily exactly satisfied during violent relaxation).
The scalings that can be extracted from these equations may be written as

$$\Gamma_{cap} \propto n_o R_d f_r/v_* \eqno{(4.1)}$$

\noindent where $\Gamma_{cap}$ is the capture rate per star, $n_o$ and $v_*$
are respectively the local stellar number density, and
velocity dispersion and $f_r$ is a reduction factor for the case in which
star-disc encounters are predominantly destructive (i.e. non-capturing):

$$f_r=1\ \ (v_*\ltaprx V_c) \eqno{(4.2)}$$

and

$$f_r \propto (V_c/V_*)^2\ \ (v_*\gtaprx V_c) \eqno{(4.3)}$$

where

$$V_c = (4GM_d/R_d)^{1/2} \eqno{(4.4)}$$

\yyskip
\noindent{\bf 4.1 Uniform Density Case}
\yyskip

\par
The results of the uniform density N-body calculations are of particular
interest since they admit detailed comparison with the analytical calculation
of MCP. As is evident from Figure 1a, the N-body results produce consistently
more binaries than the MCP calculation. This effect can be readily understood
by noting that the MCP calculations are terminated when the velocity
perturbations become non-linear (i.e. during the `bounce' phase), whereas the
N-body results can be pursued over a number of crossing times. At low N,
a number of binaries are formed post-bounce. As N is increased (for constant
$R_d$ and cluster half mass radius, $R_h$) the importance of post-bounce
encounters decreases considerably
and this progressively improves the agreement between the two models. This
effect is illustrated in Figure 9 by the $\times$ (denoted 1-7 for
progressively
higher N) whereas the open circle contains the predictions of the MCP method
for these models. Quantitatively, one may understand the steep decline in
binary formation efficiency for $N \gtaprx 2000$ by noting that star-disc
encounters start to become predominantly destructive (as opposed to capturing)
once the velocity dispersion exceeds $V_c$ (equation (4.4)). For the
values of $R_h$, $M_d$ and $R_d$ of our models, the post-bounce velocity
dispersion
exceeds $V_c$ for $N \gtaprx 2000$ and thus one would anticipate the
shutting off of post-bounce captures in clusters larger than this.
\par
We deduce from this that the MCP results give a reasonable prediction of the
binary fraction for uniform clusters of $\sim 10^4$ stars, providing estimates
that are less than a factor two below the `true' N-body results in this
regime. This success then leads us to enquire more deeply as to the way the
MCP results scale with model parameters, an investigation that is
computationally prohibitive with the N-body code at such high N.
\par
The solid dots in Figure 9 show the variation of binary fraction, $f_{bin}$,
(by the MCP method) as
a function of the initial kinetic energy parameter, $Q$, for a variety of
values of $N$ in the range $10^4$ to $10^6$, whilst the solid line is a fit
to these points of the form $f_{bin} \propto Q^{-1/2}$. Such a scaling may be
understood by considering the expressions for the capture rate per star as a
function of local variables (equations (4.1) to (4.4) above).
One crude estimate of
the resultant binary fraction may be obtained by multiplying the
initial $\Gamma_{cap}$
by the cluster free-fall time,  so that, noting that initially $v_* \propto
(QN/R_h)^{1/2}$ we obtain $f_{bin} \propto (R_d/R_h)Q^{-1/2}$ (if the initial
velocity dispersion is less than $V_c$: $Q \ltaprx  Q_c$) and $f_{bin} \propto
M_d/(NQ^{3/2})$ for $Q \gtaprx Q_c$. The former scaling, with binary fraction
independent of N, is reproduced by the solid points in Figure 9; for higher
N, $Q$ can become greater than $Q_c$ and the diamonds in Figure 9 ($N=10^6$)
indicates the $Q^{-3/2}$ scaling (dashed line) in this case.

At the low $Q$ end, as well, the binary fraction deviates from the
$Q^{-1/2}$ law, since the velocity dispersion cannot be held close to
its initial value for a cluster collapse time if this initial value is
much smaller than the initial two-body free-fall velocity. If such is the
case, then two-body effects drive up the velocity dispersion on less than
a cluster collapse time, and the resultant binary fraction is accordingly
lower than an extrapolation of the $Q^{-1/2}$ scaling; the limiting Q value
in this case is $Q_{min} \sim N^{-2/3}$. This effect is illustrated by the $+$
in Figure 9 for the case $N=10^4$. However, in the case of
a stellar cluster in which the initial stellar separation is of order a
Jeans length, this limiting initial velocity dispersion is of order the sound
speed: in reality, pressure differentials during fragmentation will always
impart a velocity dispersion of at least this order, so that the $Q \ltaprx
Q_{min}$ case would not be encountered in practice.

To summarise the results above, the resultant binary fraction in the case of
a uniform high N cluster can be expressed by the following semi-empirical
formulae:
$$ f_{bin}=3\% (R_d/10^{-3}R_h)(Q/10^{-2})^{-1/2}\ \ (Q_{min}\ltaprx Q
\ltaprx Q_c) \eqno{(4.1.1)}$$

$$ f_{bin}=1\%(M_d/M_*)(Q/10^{-2})^{-3/2}(N/10^6)^{-1}\ \ (Q\gtaprx Q_c)
\eqno {(4.1.2)}$$
where
$$ Q_{min} \sim N^{-2/3} \eqno{(4.1.3)}$$
and
$$ Q_c\sim 6\times 10^{-3}(M_d/M_*)(10^{-3}R_h/R_d)(N/10^6)^{-1} \eqno {(4.1.4)
}.$$

In each case the scalings have been derived using the arguments above, whilst
the coefficients have been fit to the results of the MCP calculations in the
range $N=10^4-10^6$ and $Q=10^{-4}-0.1$. We note, in passing, that it is
remarkable how well these scalings work (based as they are on {\it initial}
capture rates) when one considers that the bulk of binary captures occur
well into the `bounce' phase. The dependence of the binary fractions on $Q$
indicates that the system in some sense `remembers' its initial conditions
during the onset of the bounce.

The above expressions refer to the total yield of binaries generated by this
process, whose semi-major axes are initially distributed  as $N_{bin}(< a)
\propto a$ for $a \ltaprx R_d$. It is likely, however, that the ultimate
distribution of separations is mainly governed by subsequent orbital evolution
of the protobinary due to gravitational interaction with remnant disc
material. For example, if as little as $10 \%$ of the binary mass remains in
circumbinary orbit following capture, then orbital energy transfer from binary
to disc can cause the binary orbit to shrink over a time scale of a few
thousand
orbital periods (Artymowicz et al 1991). For binaries with separations
$\sim 100$ A.U., substantial orbital shrinkage could then occur before the
dispersal of the disc on a time scale $10^6-10^7$ years (Skrutskie et al 1991).
If, however, such spiralling
in does not occur during the disc lifetime, many of the binaries produced
above would be destroyed, being wider than the hard-soft borderline
$a_{hs} \sim 0.4 R_{pc} (10^6/N)$ A.U., where $R_{pc}$ is the cluster half
mass radius in parsecs.. The relevant destruction time scale is
$\sim(R_h/a)$ cluster crossing times (Binney \& Tremaine 1987) and is therefore
considerably longer than the lifetime of protostellar discs. Thus we conclude
that either substantial orbital evolution occurs during the pre-main
sequence stage (e.g. efficient binary-disc coupling), in which case $f_{bin}$
above represents the ultimate binary fraction, or else only a fraction
$a_{hs}/R_d$ of these binaries survive. In the latter case the ultimate
binary fraction (for $Q \ltaprx Q_c$) would be reduced to
$0.6 \% (Q/10^{-2})^{-1/2}(10^4/N)$.

\yyskip
\noindent{\bf 4.2 Isothermal Case.}
\yyskip

In the case of clusters with an initially isothermal density profile, we have
no model for the growth of density and velocity perturbations, and thus cannot
undertake the type of comparison described above. We can however gain some
insights into the binary formation process by consideration of Figures 3 and 4.
First, it is clear from comparison of these and Figure 1a) that the binary
formation history of isothermal systems is not characterised by a well-defined
peak, since the spread of arrival times at the origin in inhomogeneous systems
does not
result in a single `bounce'. Instead, the innermost regions of the cluster
collapse first, and outer shells, collapsing later, interact during their
infall with initially inward lying material now in the process of re-expansion
from the origin. As a result, the outer parts of the cluster acquire a velocity
dispersion that is a substantial fraction of virial during their infall phase.
This effect is apparent in both Figures 3 and 4, from which we can distinguish
two phases of binary formation. For the first $\sim 0.2$ of
a cluster free fall time scale (`early' phase) the binaries originate from the
innermost
regions which are collapsing in a manner similar to a homogeneous system (ALP):
during this time, therefore, the parts of the cluster that form most of the
binaries `remember' the initial conditions, an effect that is apparent from
the $\sim Q^{-1/2}$ dependence of the binary yield during this phase
(note also that the `early' binary fraction is insensitive to N: cf
equation (4.1.1) above). Subsequently, however, (`late' phase)  the number of
binaries formed is more or less independent of $Q$, an effect that
reflects the erasing of initial conditions through shell crossing in the
outer parts of the cluster. As $N$ is increased beyond $\approx 2000$, where
the virial velocity becomes comparable with $V_c$, the relative importance
of `late', as compared with `early', captures decreases, due to the
increasing predominance of destructive encounters in the `late' regime,
and thus the $Q$
dependence is better preserved in the total binary yields at higher $N$
(models 10 and 11 cf 12 and 13).

Despite the very different histories of binary formation in uniform and
isothermal models, it is notable how insensitive is the total binary
yield to gross changes in the global density profile.  Comparison of models
1-4, 6 and 7 with models 8-13 indicates that at comparable
times the binary yields in the two cases differ by less than a factor two,
the tendency being for isothermal models to produce somewhat fewer binaries
than corresponding uniform models.

\yyskip
\noindent{\bf 4.3 Clumpy clusters}
\yyskip

The existence of hierarchical clustering in the initial conditions changes
both the character of the collapse and the resultant binary yield as compared
to the
case of smooth initial conditions. In a system consisting of $N_c$ clumps,
with initial filling factor $f_v$, the clumps collapse on themselves on a
time scale equal to $f_v^{1/2}$ times the cluster free fall time scale. If
these time scales are separated by a factor of more than a few, the clumps
will have undergone violent relaxation (and re-expanded to a dimension
$\sim$half their
initial sizes) by the time the whole cluster reaches maximum compression.
Since the maximum collapse factor of the whole system is $\sim N_c^{1/3}$, the
clumps will not merge at the bounce in
any cluster for which $f_v \ltaprx 8/N_c$. Consequently, such systems (such
as model 14) evolve as an ensemble of $N_c$ independent clumps, an expectation
that we have confirmed by comparing the binary yield in this case with that
from a single such clump (model 17). The binary fraction may therefore be
increased in
clumpy models due to two effects: a) the reduced system size and consequently
increased density and b) if $N/R$ is smaller for each clump than for the
cluster as a whole, the lowered internal velocity dispersion reduces the
incidence of fast, non-capturing encounters. From models 15, 16 and 1-5 in
Table I (identical models apart from progressively greater $N$ in the
range $50$ to $10^4$) we deduce that the binary yield increases with increasing
$N$ (because of increased density) until such point that the virial velocity
becomes comparable with $V_c$; for the disc to cluster radius of these models,
the optimum binary yield is obtained for clusters numbering one to two
thousand (Figure 9).

\yyskip
\noindent{\bf 4.4 Dependence on Disc Parameters}
\yyskip

We now consider the dependence of binary yield on disc parameters implied by
equations (4.1)-(4.4). In the case where the velocity dispersion is below
$\sim V_c$
the determining factor is $R_d$ alone, since most encounters that intersect
the disc result in capture. Conversely, where the velocity dispersion exceeds
$V_c$,
only a fraction of star-disc encounters result in capture: mainly those that
hit the disc within a radius well inside $R_d$, this critical radius depending
on the strength of star-disc interaction, i.e. $M_d$.

For the case of clusters
that remain in the former regime for much of a free-fall time, the chief
disc quantity that determines the binary yield is $R_d$, the initial value of
which being fixed by the angular momentum retained, or acquired, during the
fragmentation process.  One scenario, assumed by MCP, comes about from an
analogy with the angular gained by protogalactic clouds.  During their
fragmentation, gravitational torques from neighboring fragments give the
condensations an angular velocity such that their energy of rotation is a
constant fraction ($\lambda$) of their gravitational binding energy.
Taking the value
$\lambda\approx0.07$ found in cosmological simulations of fragmentation
(Layzer 1963; Barnes \& Efstathiou 1987) and assuming that the initial
fragment dimensions are $\sim$ the initial interstellar separation, we thus
obtain the scaling
$$R_d/R_h\sim10^{-3} (N/10^6)^{-1/3}(\lambda/0.07).\eqno{4.4.1}$$

It may
plausibly be argued, however, that such a picture is only appropriate in the
case that fragments have had sufficient time to redistribute
angular momentum in this way and that this may not be the case in the
`prompt initial fragmentation' scenario envisaged here.  If this is so, then
the angular momentum per fragment is either gained as a result of the star
formation process of cloud-cloud collision followed by thermal and
gravitational instability, or that which it inherits from the rotation of the
gaseous protocluster.  The angular momentum expected to result in the former
case is uncertain, but will almost certainly not be less than that expected
in the latter case, which can therefore be treated as a lower limit.  Provided
that the initial cluster rotation is fixed by $\lambda$, due to encounters
during the fragmentation of the protocluster clouds, then the resultant mean
disc radius is again similar to that derived above (the reason for this may be
understood by noting that as the fragments condense out of the background,
their densities are comparable with the mean cluster density and thus the
break-up  angular velocity of cluster and of individual fragments are similar).
While these different processes lead to similar disk radii, they do have
different consequences
for the way that disc radii scale with stellar mass (see \S\ 4.5 below)
and also because the latter scenario ties the disc radii more directly to the
rotation of the parent cluster.  Variations in binary fractions would, in this
case, be related to the angular momentum imparted to the cluster during
its initial fragmentation.  Such a scenario could, in principle, be tested
observationally, though it would be extremely difficult, given the small
rotational energies expected (and observed) in clusters.

In the numerical calculations we have taken $R_d =5\times 10^{-4} R_h$ in most
of our models. Such a value is comparable to that implied by equation (4.4.1)
for higher $N$ models.  It is an order of magnitude greater than that employed
by MCP, due to the higher densities for the star forming gas assumed in the
previous work.  Such differences should be taken as reflecting uncertainties
in disk radii expected to result from the star formation process.  We adopt
the larger values throughout this study in order to reduce the computational
inaccuracies involved in calculating very close encounters.

We stress that though any value of
$R_d$ may be linked, by the above arguments, to a cluster rotation rate,
all our simulations have been undertaken with zero angular momentum
clusters. We note that since a rotating cluster collapses by a factor
$\lambda^2$ before being held up by centrifugal forces, and since a
uniform cluster collapses by a factor $N^{1/3}$ before entering the
`bounce', rotation is dynamically unimportant for all clusters for
which $N \ltaprx 10^7(0.07/\lambda)^6$.  A link between disk radii
and angular momentum of the parent cluster does not, therefore, conflict
with the observed lack of rotation in globular clusters.

We now turn to the case in which $M_d$ and $R_d$ undergo viscous evolution
during the cluster collapse. Here, our N-body results show a remarkable
insensitivity to such evolution on a time scale comparable with the cluster
free-fall time scale. This result can be traced to the fact that angular
momentum conservation requires $M_d$ evolves as $R_d^{-1/2}$ (for {\it any}
viscosity prescription) so that the resultant capture rates scale as
$\Gamma_{cap} \propto R_d$ and $\Gamma_{cap} \propto R_d^{-1/2}$ in the low
velocity and high velocity regimes respectively. Thus we find that although
viscous evolution boosts the capture rate somewhat at early times (slightly
larger discs) this is offset by the reduced disc mass at late times (smaller
fraction of capturing encounters). This insensitivity to viscous evolution
results from the relatively large values of the initial disc radii used
in our N-body calculations, which mean that the crossover from
$v_* < V_c$ (equation 4.4) to $v_* > V_c$ occurs relatively early in the
collapse. For smaller disc initial radii (as employed by MCP), viscous
evolution can boost the binary yield by a factor of a few.

\yyskip
\noindent{\bf 4.5 Dependence on Stellar Mass Function.}
\yyskip

It is well known that in the process of violent relaxation, stars acquire a
specific energy that is independent of their mass, thus behaving, in this
respect, like test particles in the rapidly varying cluster potential. It
is therefore no surprise that the global properties of the collapse are little
affected by the introduction of a spread of stellar masses in models 20-22.

In this Section we consider the mass dependence of the binary formation
process: specifically, we derive scalings for the distribution
of companion masses, $M_2$, to stars of (fixed) mass $M_1$. To clarify the
argument we consider here only the case $M_2 < < M_1$, since in this case
the degree of gravitational focusing of an encounter (for given boundary
conditions at infinity) is $\sim$ independent of $M_2$. We note, from the
property of violent relaxation alluded to above, that the velocity dispersion
is independent of mass at all times, and that (owing to the lack of mass
segregation) the mass function also remains spatially uniform at all times. It
follows, therefore, that any deviation of the binary pairing process from
random selection from the mass function must result from the nature of
the star-disc interaction process: from the dissipative prescription
employed and from the dependence of disc mass and radii on stellar mass.

In the numerical simulations, and in the scalings below,
we employ the interaction prescription
equation (2.1.2) and also the scalings $M_d \propto M_*$ and $R_d \propto M_*$.
The disc mass scaling results from the (not unreasonable) assumption that
the initial division of fragment mass into star and disc is scale free. The
disc radius scaling results from assuming a scenario in which condensations
are spun up to a constant fraction of break up during fragmentation (see
discussion in \S\ 4.4 above).
If such is the case, fragments collapse by a constant factor
from dimensions $\sim$ a Jeans length, $R_J$. Since, for fluctuations of
various densities in an isothermal medium, $R_J \propto M_J$ (the Jeans mass)
it follows that in this case $R_d \propto M_*$. We note, however, that if we
adopt the alternative view (i.e. that the angular momentum of
condensations is inherited from the rotation of the parent cluster: see
\S\ 4.4) the resultant scaling would be $R_d \propto M_*^3$. We
mention this point in order that the uncertainties entering
these scalings can be appreciated.

The prescription for energy loss on star-disc interaction (equation 2.1.2)
implies

$$\Delta E \propto M_{d1}M_2^2 + M_{d2}M_1^2 \eqno{(4.5.1)}$$

whereas the initial energy of the relative orbit $E_{orb} \propto M_2$.
{\it If} $M_d \propto M_*$ (see above) then it follows that $\Delta E/E_{orb}
\propto M_2$ if only $M_1$ has a disc whereas $\Delta E/E_{orb}$ is
independent
of $M_2$ if both stars have a disc or if $M_2$ only has a disc. If $M_1$
has a disc (whether or not $M_2$ has a disc) it follows that it is the
dimension of $M_1$'s disc that mainly fixes whether an encounter takes
place (since $R_d \propto M_*$), whereas if only $M_2$ has a disc it is
clearly the size of its disc that is the determinant. Putting these effects
together we conclude that if both stars have discs the process is $\sim$
independent of $M_2$ (i.e. random picking from the mass function) whereas if
only one star has a disc the capture process is biased against companions
of low $M_2$. We can therefore understand the downturn in the companion mass
function in Figure 8b and 8c as  resulting both from the form of the energy
loss prescription used and from the fact that lighter stars have smaller discs,
and are thus less likely to be involved in star-disc encounters.

\yyskip
\noindent{\bf 5. CONCLUSIONS }
\yyskip

\par
We have used N-body simulations of globular cluster collapse
in order to clarify what are the major factors that determine the
efficiency of binary formation through star-disc captures.

We find that in the case of large $N$ clusters with a smooth density profile,
the  main determinants are the ratio of disc to cluster radii ($R_d/R$) and
the initial stellar velocity dispersion (equation 4.1.1).
Both these quantities are
determined by the details of the fragmentation process, and the former also
by the rotation of the protoglobular cloud. The maximum binary yield is
obtained when the initial stellar velocity dispersion is as low as possible,
that is of order the
sound speed in the star forming gas. Using disc radii implied by the
rotation rates estimated for protoglobulars from cosmological simulations,
we obtain a maximum binary yield of $3\%$ in such systems, independent
of $N$ (equations 4.1.1, 4.1.3 and 4.4.1). It is notable
that this maximum yield scales as $\lambda$ where $\lambda$ is the ratio of
rotational to gravitational energy of the protostellar condensation
(assumed to be 0.07 in the estimate above).  If, instead of resulting from
the process of star formation, the angular momentum of the protostars is
inherited from the parent cluster, then we expect that more rapidly rotating
clusters should yield a higher binary fraction.

The effect of viscous disc evolution (on a time scale comparable with
the cluster collapse time scale) is to cause discs simultaneously to
grow in size (binary promoting) and to shrink in mass (binary inhibiting).
Our numerical simulations use rather large initial disc radii, for
reasons of computational economy, and in this case further viscous
growth hardly affects the binary yield: the effects of the above
two processes roughly cancel. In previous (analytical) work using
smaller initial disc sizes we showed that viscous evolution can boost the
binary yield calculated above by a factor of a few (MCP).

Our N-body simulations (with  $N$ as high as $5000$) show that the binary
yield is remarkably insensitive both to changes in the initial stellar
density profile and to the inclusion of a range of stellar masses.
Changing the initial density distribution from uniform to that of
an isothermal sphere changes the nature of the collapse, since, in the
latter case, the inner regions collapse first and, in re-expanding,
interact with material infalling from larger radii. As a result, the history
of binary formation is no longer marked by a well defined peak in production
at maximum compression (the `bounce'). Remarkably, however, the binary
yields are rather similar in the two cases, with comparable uniform
models exceeding the isothermal models by less than a factor two.

The insensitivity
of over-all binary yields to the inclusion of a spectrum of stellar masses
may be understood by noting that, in the process of violent relaxation, the
specific energy attained by each particle is independent of mass. The mass
dependence of the binary pairing process is therefore governed only by the
way that the star-disc energy loss prescription and the masses and radii
of discs scale with
stellar mass. Such details, which are not well understood theoretically at
present, will determine the precise form of the binary statistics as a
function of mass. We note, however, that since it is a reasonable expectation
that more massive stars possess larger discs, it is likely that the binary
pairing process is more biased toward massive stars than would result from
random pairing from the mass function (Figure 8).

We stress that all the binary yields quoted here are total binary yields,
without reference to whether they are hard or soft and are thus likely to
survive in the cluster environment. In fact, since most star-disc capture
binaries are formed with orbital velocities comparable with the local
velocity dispersion at the time of formation, and since, in high $N$
clusters, most binaries are formed when the velocity dispersion is
sub-virial, it follows that most of the binaries formed in this way are
soft. If one assumes that all initially soft binaries are destined to
be dissolved by encounters with field stars, then the resultant yields
of surviving binaries are reduced considerably, to $\ltaprx1\%$.
Calculations of the interaction between  a pre-main sequence
binary and material in circumbinary orbit however indicate that binary
orbits can be efficiently shrunk during the pre-main sequence stage,
even where the mass of gas left in circumbinary orbit is a rather
modest fraction of the binary mass ($\sim10\%$; Artymowicz et al 1991).
If such is the case, then the bulk of binaries formed through star-disc
captures could survive, although naturally, in this case, their period
distributions would then reflect this subsequent orbital evolution, rather
than the initial capture parameters.

Finally we note  that binary yields considerably higher than those quoted
above for smooth initial conditions can be obtained if the initial
stellar distribution is clumpy.
In this case, overdense regions, collapsing ahead
of the general cluster, may cause the system to behave like an ensemble of
independent small $N$ systems over a number of dynamical time scales. This
may considerably increase the yield, both because the virial velocity of
each clump may be less than that of the cluster as a whole, implying a
preponderance of capturing (as opposed to disc destructive encounters) over
many dynamical times, and also because of the enhanced density within each
clump. For example, we find that by dividing a cluster into an ensemble of
stellar `nests', each containing several hundred stars and several tenths
of a parsec in radius, we obtain binary fractions of $\gtaprx 15\%$. Such
lumpiness in initial conditions would persist over several cluster crossing
times, and would be compatible with the structure seen in the Magellanic
globulars of about that age.

\yyskip
The authors gratefully acknowledge helpful conversations with S. Aarseth
and J. Pringle.  S. D. M. was supported during this work by a NATO Postdoctoral
Fellowship awarded in 1991.

\par\vfill\eject
\centerline{\bf References}
\yyskip

\pp Aarseth, S. J. 1985, in {\it Multiple Time Scales}, ed. J. U. Brackbill,
and B. I. Cohen (Orlando: Acad. Press), p. 377

\pp Aarseth, S. J., Lin, D. N. C., \& Papaloizou, J. C. B. 1988, ApJ, 324,
288 (ALP)

\pp Artymowicz, P., Clarke, C.J., Lubow, S.H. \& Pringle, J.E. 1991, ApJL
370, L35.

\pp Barnes, J. \& Efstathiou, G. 1987, ApJ, 319,575

\pp Boily, C., Clarke, C.J. \& Murray, S.D. 1992, in {\it Proc 11th Santa
Cruz Workshop on Astronomy and Astrophysics}, ed. G. Smith, in press.

\pp Bolte, M. 1987a, ApJ, 315, 469

\pp ----- 1987b, ApJ, 319, 760

\pp Clarke, C. J., \& Pringle, J. E. 1991a, MNRAS, 249, 584 (CP)

\pp ----- 1991b, MNRAS, 249, 588

\pp Cohen, J. G. 1979, ApJ, 231, 751

\pp Elson, R. A. W. 1991, ApJS, 76, 185

\pp Fabian, A.C., Pringle, J.E. \& Rees, M.J. 1975, MNRAS, 172, 15P

\pp Gao, B., Goodman, J., Cohn, H. N. \& Murphy, B. 1991, ApJ, 370, 567

\pp Goodman, J. 1989 in {\it The Dynamics of Dense Stellar Systems} ed. D.
Merritt (Cambridge: Cambridge University Press) p. 183

\pp Goodman, J. \& Hut, P. 1989 Nature 339,40

\pp Hills, J.G. 1976, MNRAS 175 1P

\pp Lada, C. J., Margulis, M., \& Dearborn, D. 1984, ApJ, 285, 141

\pp Larson, R.B. 1990 in {\it Physical Processes in Fragmentation and Star
Formation}, eds R. Capuzzo-Dolcetta, C. Chiosi \& A. di Fazio, (Dordrecht:
Kluwer Academic Publishers), p. 389

\pp Layzer, D. 1963, ApJ 137,351.

\pp Lin, D. N. C., \& Pringle, J. E. 1990, ApJ, 358, 515

\pp McMillan, S.L.W., Hut, P. \& Makino, J. 1990, ApJ 362, 522

\pp------- 1991 ApJ 372,111

\pp McMillan, S.L.W., McDermott, P.N. \& Taam, R.E. 1987 ApJ 318,261

\pp Murray, S.D., Clarke, C.J. \& Pringle, J.E. 1991 ApJ 383,192 (MCP)

\pp Murray, S.D. \& Lin, D.N.C. 1989 ApJ 339,933

\pp Pryor, C.P., Latham, D.W. \& Hazen, M.L. 1988, AJ 96,123

\pp Pryor, C.P., Schommer, R.A. \& Olszewski E. W., 1991 in {\it
The Formation and Evolution of Star Clusters}, ed K. Janes, (San Francisco:
Astron. Soc. Pacific), p. 439

\pp Pringle, J.E. 1989, MNRAS, 239,361

\pp Richer, H. B., \& Fahlman, G. G. 1984, ApJ, 277, 227

\pp Romani, R.W. \& Weinberg, M.D. 1991 ApJ 372,487

\pp Sandage, A., \& Katem, B. 1977, ApJ, 215, 62

\pp ----- 1982, AJ, 87, 537

\pp Shu, F. H., Adams, F. C., \& Lizano, S. 1987, ARA\&A, 25, 23

\pp Tenorio-Tagle, G., Bodenheimer, P. H., Lin, D. N. C., \& Noriega-Crespo, A.
1986, MNRAS, 221, 127

\par\vfill\eject

\singlespace
\def\pt{\par\yskip\noindent\hangindent 0.25in \hangafter 1}

$$
\vbox{%
\noindent{\bf Table 1.}  The Models.
\vskip 10pt
{\halign {\hfil # & \quad \hfil # & \quad \hfil #\hfil & \quad \hfil #\hfil
& \quad \hfil #\hfil & \quad \hfil # & \quad \hfil #\hfil & \quad \hfil # &
\quad \hfil # \cr
Model & N$_\ast$ & $\rho_\ast(R){^a}$ & Q & Evolution$^b$ & R$_d$\ (pc) &
IMF$^c$ & N$_{bin}$ & $\tau_{bin}^d$ \cr
   &      &   &      &     &                  &   &     &     \cr
1  &  500 & u & 0.05 &  no &        10$^{-3}$ & U &  26 & 2.0 \cr
2  & 1000 & u & 0.05 &  no &        10$^{-3}$ & U &  70 & 2.0 \cr
3  & 2000 & u & 0.05 &  no &        10$^{-3}$ & U & 122 & 2.0 \cr
4  & 5000 & u & 0.05 &  no &        10$^{-3}$ & U & 112 & 2.0 \cr
5  &10000 & u & 0.05 &  no &        10$^{-3}$ & U & 141 & 2.0 \cr
6  & 2000 & u & 0.01 &  no &        10$^{-3}$ & U & 134 & 1.3 \cr
7  & 5000 & u & 0.01 &  no &        10$^{-3}$ & U & 203 & 1.0 \cr
8  &  500 & i & 0.05 &  no &        10$^{-3}$ & U &  25 & 1.0 \cr
9  & 1000 & i & 0.05 &  no &        10$^{-3}$ & U &  44 & 1.0 \cr
10 & 2000 & i & 0.05 &  no &        10$^{-3}$ & U &  66 & 1.5 \cr
11 & 5000 & i & 0.05 &  no &        10$^{-3}$ & U &  72 & 1.3 \cr
12 & 2000 & i & 0.01 &  no &        10$^{-3}$ & U &  87 & 2.0 \cr
13 & 5000 & i & 0.01 &  no &        10$^{-3}$ & U & 121 & 2.0 \cr
14 & 2000 & c & 0.01 &  no &        10$^{-3}$ & U & 317 & 2.0 \cr
15 &  100 & u & 0.01 &  no &        10$^{-3}$ & U &   5 & 2.0 \cr
16 &   50 & u & 0.01 &  no &        10$^{-3}$ & U &   1 & 2.0 \cr
17 &  200 & u & 0.01 &  no & 5$\times10^{-3}$ & U &  39 &10.0 \cr
18 & 2000 & u & 0.01 & yes &        10$^{-3}$ & U & 138 & 1.3 \cr
19 & 2000 & i & 0.01 & yes &        10$^{-3}$ & U &  91 & 2.0 \cr
20 & 2000 & u & 0.01 & yes &        10$^{-3}$ & S & 134 & 2.0 \cr
21 & 2000 & u & 0.01 &  no &        10$^{-3}$ & S & 134 & 2.0 \cr
22 & 2000 & i & 0.01 & yes &        10$^{-3}$ & S &  68 & 2.0 \cr
}}}
$$

\pt{$^a$}u = uniform distribution, i = $\rho_\ast\propto R^{-2}$, c =
clumpy distribution (see text).

\pt{$^b$}no = no disk evolution, yes = disks evolve as described in text.

\pt{$^c$}U = single stellar mass, S = Salpeter IMF.

\pt{$^d$}Time of evaluation of N$_{bin}$ in units of cluster crossing time.

\vfill\eject

\oneonehalfspace
\centerline{\bf Figure Captions}

\noindent{\bf Figure 1.}  (a) $N_{bin}$ vs. time for models 1-5 (solid, dotted,
short dashed, long dashed, dot-dashed curves respectively).
The crosses indicate the values
of $N_{bin}$ and time when the models of MCP became nonlinear. (b)
histograms of the resulting
semimajor axis distributions for models 1 and 5 (solid and short dashed).
The vertical tick marks
indicate, from left to right, the softening lengths used in the models 5 and
1 and the disk radii.

\noindent{\bf Figure 2.}  The number of binaries (solid curve) and encounters
which do not lead to capture (short dashed curve) for model 5.

\noindent{\bf Figure 3.}  $N_{bin}$ vs time for $N=2000$ systems
with unevolving disks: models 3 (solid), 6 (dotted), 10 (short dashed)
and 12 (long dashed).

\noindent{\bf Figure 4.} $N_{bin}$ vs time for $N=5000$ isothermal
systems: models 11 (solid) and 13 (long dashed).

\noindent{\bf Figure 5.}  $N_{bin}$ vs time for uniform and clumpy systems:
models 12 (short dashed) and 15 (clumpy).

\noindent{\bf Figure 6.}  (a) $N_{bin}$ vs time for models 6 and 18
(short dashed and solid) and $N_{loss}$ vs time in each case (long dashed
and dotted). (b) Histograms of resulting semi-major axis distribution
(short dashed and solid respectively).

\noindent{\bf Figure 7.}(a)  As Figure~6a for models 12 (short dashed and
long dashed) and 14 (solid and dotted). (b) Histograms of resulting
semi-major axis distribution (short dashed and solid respectively)

\noindent{\bf Figure 8.} (a) Fraction of stars in binaries as a function
of stellar mass for models 20 (solid) and 22 (short dashed). (b)
Mass ratio distribution for binaries containing one member in range
$0.7-1.3 M_{\odot}$ for model 20 (solid) compared with  distribution
expected in the case of random picking from the mass function (dotted). (c) as
(b) for model 22.

\noindent {\bf Figure 9.} Binary fraction as a function of Q for clusters
with $R_d/R = 5 \times 10^{-4}$. Crosses (1-7) are N-body results for
successively higher $N$ in the range $50$ to $10^4$. Other points are
calculated
by MCP method: solid points corresponding to equation (4.1.1) (solid line),
diamonds (for $N=10^6$) to equation (4.1.2) (dotted line) and crosses to
the regime $Q < Q_{min}$ (equation 4.1.3) for $N=10^4$.

\par\vfill\eject
\end